# Battery Asset Management with Cycle Life Prognosis

Xinyang Liu, Pingfeng Wang, Esra Büyüktahtakın Toy and Zhi Zhou

*Abstract*—Battery Asset Management problem determines the minimum cost replacement schedules for each individual asset in a group of battery assets that operate in parallel. Battery cycle life varies under different operating conditions including temperature, depth of discharge, charge rate, etc., and a battery deteriorates due to usage, which cannot be handled by current asset management models. This paper presents battery cycle life prognosis and its integration with parallel asset management to reduce lifecycle cost of the Battery Energy Storage System (BESS). A nonlinear capacity fade model is incorporated in the parallel asset management model to update battery capacity. Parametric studies have been conducted to explore the influence of different model inputs (e.g. usage rate, unit battery capacity, operating condition and periodical demand) for a five-year time horizon. Experiment results verify the reasonableness of this new framework and suggest that the increase in battery lifetime leads to decrease in lifecycle cost.

## I. Introduction

IN the past three decades, the applications for lithium-ion batteries as major energy storage devices have spread into an increasing number of fields related to human life, such as smart phones, personal computers and electric vehicles. Moreover, energy storage technologies are expected to play a decisive role in the future development of renewable energy systems [1], considering an increasing penetration of renewable energy-based power generation units (e.g. solar and wind) in a modern power grid. Battery energy storage system (BESS), provided with the maturity of battery technology and its operation management, could substantially enhance the reliability and resilience of critical infrastructure systems, such as power transmission and distribution systems [2]. Suitable battery choices and advanced technologies applied to the BESS have been discussed extensively in the literature [3, 4]. With the prevalence of energy storage installations at the utility scale, BESS assets gradually become a new important type of assets for power systems asset owners. To achieve high operational reliability and functionality robustness while increasing the profitability of the BESS assets, it is imperative to develop an asset management platform with technical tools for the BESS asset owners to manage their assets better. The asset management platform should be able to take into account special characteristics that BESS assets are different from other types of assets and enable operational cost optimization for a given period of designed service life [2].

As unexpected battery failures could result in enormous economic and societal losses, safe and reliable operation of lithium-ion batteries is of vital importance. Extensive research has been performed in the past decade for the development of effective battery management systems, and good reviews of these developments can be found from the literature [5, 6]. Because capacity fade and internal resistance increase due to aging of battery cells, directly affecting the performance of a battery pack by decreasing both energy and power outputs, two important parameters: state-of-charge (SoC) and state-of-health (SoH) are applied to indicate battery health conditions. For accurate assessment of the performance of an operating battery cell, a number of techniques for SoC and SoH estimations have been developed in the literature [7]. One of the most commonly used SoC estimation approaches is the ampere hour counting technique [8], which calculates SoC values by integrating current with respect to time. Due to its high accuracy, the ampere hour counting technique has been used primarily as a benchmark method in the research community. In common practice, battery manufacturers generally utilize open circuit voltage (OCV) measurements to find out corresponding SoC values from SoC-OCV tables, which are expensively made based on experiments by comparing SoC and OCV under different operating conditions [9]. To avoid extensive efforts in developing the SoC-OCV tables, advanced battery power management techniques have also been developed recently [10-16]. For example, He et al. developed an approach using Dempster-Shafer theory (DST) and the Bayesian Monte Carlo (BMC) method for the estimation of both SoH and remaining useful life (RUL) [17]. A self-cognizant dynamic system-based prognosis approach has been developed by Bai and co-workers and applied to battery RUL prediction [18]. Hu et al. developed an approach to estimate the SoH and predict the RUL using the

Manuscript received August, 2020. This work was partially supported by National Science Foundation through Faculty Early Career Development (CAREER) awards: CMMI-1351414 (P. Wang) and CBET-1554018 (E. Büyüktahtakın Toy).

Xinyang Liu and Pingfeng Wang are with the Department of Industrial and Enterprise Systems Engineering, University of Illinois at Urbana Champaign, Urbana, IL 61801 USA (e-mail: xl50@illinois.edu, pingfeng@illinois.edu).

Esra Büyüktahtakın Toy is with the Mechanical and Industrial Engineering Department, New Jersey Institute of Technology, Newark, NJ 07102 USA (e-mail: esratoy@njit.edu).

Zhi Zhou is with the Argonne National Laboratory, Lemont, IL 60439 USA (email: zzhou@anl.gov)

Gauss–Hermite particle filter technique [19]. A model-based dynamic multi-parameter method was proposed to estimate the peak power of Li-ion batteries by Sun et al. [20]. Waag et al. investigated the battery impedance characteristics at different conditions and demonstrated the decreasing of SoC range due to significant aging when operating with high efficiency [21].

While battery power management studies have been conducted at battery cell and module levels, management of BESSs as physical assets has barely been investigated, primarily because of the fact that the deployment of large scale BESSs has only become prevalent in recent years. Asset management is a systematic process of developing, operating, maintaining, upgrading, and disposing of assets in the most cost-effective manner, including all costs, risks and performance attributes [22, 23]. Asset management coordinates the financial, operational, maintenance, risk, and other asset-related activities of an organization to realize more value from its assets, and over the past several decades the research community has accumulated a large number of diverse asset replacement models and methods. Yatsenko and Hritonenko [24] provided a good summary and categorization of the literature on the asset replacement models. In their study, the asset replacement models have been classified in accordance with their specific features as: a) series replacement and parallel replacement models, b) discrete-time and continuous-time replacement models, c) deterministic and stochastic models, d) models with constant and variable lifetime of assets, e) models with finite and infinite forecast horizons, and f) models with continuous and discontinuous technological change. One of the asset replacement models in particular considered in this study is the parallel asset replacement model, which determines the minimum cost replacement schedule for each individual asset in a group of assets that operate in parallel and are economically interdependent due to the fixed cost of replacement [25, 26]. In this model, the replacement of assets is often affected by increased operating and maintenance costs of deteriorating assets, or the availability of newer, more efficient assets in the marketplace. Unlike serial (single asset) replacement problems, parallel replacement problems are combinatorial as groups of assets must be analyzed simultaneously under a fixed-cost replacement. The combinatorial nature of the problem makes it NP-Hard, a very difficult problem to optimize, as proven in the study of Büyüktahtakın et al. [26].

Comprehensive studies summarizing outstanding parallel asset management policies have also been reported in the literature [26-29] and the model presented by Büyüktahtakın and Hartman [27] is employed and further modified in this study for the battery assessment management. A battery asset has unique life characteristics as its cycle life varies under different operating conditions and capacity decreases due to usage. The significance of battery cycle life prediction has resulted in a tremendous amount of research developments in this field, leading to advanced battery cycle life prognosis and power management techniques. Current life prediction models for batteries can be divided into three categories [30-32]: mechanism models, semi-empirical models and empirical models. Among all the models, stress factors, such as temperature, depth of discharge (DOD), and charge rate are employed most commonly and thus these factors are also selected in this study.

This study employs a parallel asset management model as a fundamental framework while incorporating battery cycle life prognosis information into the battery asset management decision making. It presents a mathematical programming model for the battery asset replacement problem for the first time, and further develops an asset replacement planning method to minimize the total lifecycle cost in battery energy storage systems. The rest of this article is organized as follows. In section II, battery lifetime prediction model and aging index of battery assets are introduced. Section III formally states the modified parallel asset management model. Section IV is dedicated to the effects of principle inputs and provides experimental results that illustrate the efficiency of our approach. Model application extensions are discussed in Section V and the final section concludes the article and provides some future directions.

## II. BATTERY LIFETIME PREDICTION

This section presents a lifetime prediction model and the aging index of battery assets that will be integrated with the parallel asset management model in Section III. Section II-A presents a three-parameter semi-empirical capacity fading model; Section II-B then introduces the modeling of battery aging process considering degradation due to both usage and calendar fading effect; Section II-C provides the battery lifetime prediction considering different operation scenarios.

### A. Nonlinear Capacity Fade Modeling

This study employs a three-parameter semi-empirical model introduced in [33] to predict the nonlinear capacity fade of Lithium-ion batteries caused by the growth of solid electrolyte interface (SEI) layer. During the battery charging process, a passive SEI layer is generated at electrode-electrolyte interface and continuously grows through the electrochemical side reactions, leading to an irreversible consumption of lithium ions.

In this study, the capacity fade due to the SEI growth is assumed to be occurred on the negative electrode during the charging process, and the capacity loss of a battery cell from its first to the $N^{th}$ cycle can be quantified by integrating the current density of this side reaction over time as (1),

$$Q_{\text{loss}}^{(N)} = S_{neg} \sum_{n=1}^{N} \int_0^{t_{cc,n}} J_S^{(n)} dt \quad (1)$$

where $S_{neg}$ represents the total interfacial area of the anode, $t_{cc,n}$ is the charge time at the constant current stage of the $n^{th}$ cycle, and $J_S^{(n)}$ denotes the side reaction current density $J_S$ for the $n^{th}$ cycle. Since the nonlinear property of capacity fade is caused by the deceleration of SEI growth as the SEI layer becomes thicker, $J_S$ can be presented as (2),

$$J_S = \beta_0 e^{-\lambda \delta} \exp(-E_a / RT_{\text{int}}) J_k \quad (2)$$

where $\beta_0$, $E_a$, $R$, $T_{\text{int}}$ represent a temperature-independent factor, the activation energy, the ideal gas constant, and the internal temperature of a battery cell, respectively; $\lambda$ is a limiting coefficient [34], $\delta$ is the thickness of the SEI film, $J_k = C_n/3600(s) \cdot C$ is the deintercalation/intercalation current density of Li-ion from/into the solid particles, $C_n$ is the nominal capacity of the battery, and $C$ is the charging rate, i.e. $1C$ charging rate means it takes 1 hour to fully charge the battery. As shown in the equation, the Arrhenius form is employed to characterize the temperature dependence and a multiplier of $e^{-\lambda\delta}$ describes the impact of SEI growth on the chemical reaction rate. Due to the fact that SEI layer grows thicker in a rising temperature [34-39], the model uses an inverse Arrhenius form of $\lambda$ as (3),

$$\lambda = \lambda_0 \exp(E_a / RT_{int}) \quad (3)$$

with $\lambda_0$ being a constant. In addition, during the charging/discharging processes, the temperatures of the batteries will change due to the heat generation in lithium-ion batteries [40]. In this study, the heat generation is mainly attributed to the Joule heat due to charge transport, and other heat sources are neglected since they have relatively small impacts [41]. The temperature profile of the battery in one cycle is simplified as follows. At first, the internal temperature $T_{int}$ is viewed equivalent as the ambient temperature. During the charging process, the temperature rises $\Delta T$ linearly with respect to time t. Based on experiment data in [40], $dT = 4C/t_{cc,n}dt$ where $1C$ charging leads to a temperature change: $\Delta T = 4K$. Afterwards, when discharging, battery gradually cools down to the ambient temperature due to the heat dissipation.

Based on the above description, this degradation model includes three temperature-independent parameters: $E_a$, $\beta_0$, and $\lambda_0$. $E_a$ is obtained from (4) by having the experimental plot of capacity versus cycle number at two different ambient temperatures.

$$\frac{Q_{loss}^{(N,T_1)}}{Q_{loss}^{(N,T_2)}} = \frac{e^{-E_a/RT_1}}{e^{-E_a/RT_2}} \quad (4)$$

And the other two parameters $\beta_0$ and $\lambda_0$ can be determined by finding the intercept and slope of the straight line, which is obtained by taking the logarithm of (2) as

$$\ln|J_S| + E_a/RT - \ln|J_k| = \ln(\beta_0) - \lambda_0 \delta e^{E_a/RT} \quad (5)$$

The parameters of the cycling capacity fade model for three types of commercial LIBs, i.e. LiFePO4 (LFP), LiNiMnCoO2 (NMC) and LiNiCoAlO2 (NCA), are summarized in Table. I.

The capacity fade model in [33] supposes that the capacity loss only occurs in the charge state and mainly considers the influence of temperature and charge rate. However, battery energy storage system may employ different depth of discharge values in real practice resulting in different charge time of each cycle. In our capacity fade prediction, we also introduce the flexibility of $t_{c,nn}$ considering the impact of depth of discharge ($DOD$).

Moreover, there exists battery relaxing time during the usage period when batteries are not being charged or discharged. The capacity of batteries may continue degrading during the relaxing time, which is known as calendar aging [42]. The battery operating temperature and state of charge ($SOC$) level play a decisive role in the calendar aging of Lithium-ion batteries. Fig. 1 illustrates the assumed $SOC$ profile of the batteries in this study, where the usage frequency $f = 3$, namely 3 charging/discharging cycles are performed in one day. The charging and discharging time during the cycling process is neglected for simplification. In each cycle (8 hrs in the case of Fig. 1), we assume that the batteries will be kept at full for half of the relaxing time (4 hrs in the case of Fig. 1) while stay at $SOC = 100\% - DOD$ in the other half. Temperature effect on the battery calendar aging under different levels of $SOC$ is obtained from [42] and fitted as (6),

$$Q_{loss}^{SOC} = \begin{cases} \alpha_1 \cdot \dfrac{SOC}{f} \cdot (\dfrac{T_{oper}}{T_0} - 1) \cdot C_n & SOC \leq 40\% \\ \alpha_2 \cdot \dfrac{1}{f} \cdot (\dfrac{T_{oper}}{T_0} - 1) \cdot C_n & 40\% < SOC \leq 70\% \\ \alpha_3 \cdot \dfrac{1}{f} \cdot (\dfrac{T_{oper}}{T_0} - 1) \cdot C_n & SOC \geq 70\% \end{cases} \quad (6)$$

where $Q_{loss}^{SOC}$ is the capacity loss due to calendar aging in each battery relaxing interval (4 hrs), $T_{oper}$ is the ambient temperature in Celsius, $\alpha_1$, $\alpha_2$, and $\alpha_3$ are the calendar aging factors, and $T_0$ is a fitted constant. Table. II lists the fitted parameters of the calendar aging model of three types of LIBs using the experimental data in [42]. Note that, the temperature effect on battery degradation under full $SOC$ $Q_{loss}^{full\ SOC}$ is considered to be the same with that when $SOC \geq 70\%$. Finally, calendar aging induced capacity loss during relaxing time is the sum of $Q_{loss}^{SOC}$ and $Q_{loss}^{full\ SOC}$.

TABLE I
CYCLING CAPACITY FADE MODEL PARAMETERS FOR THE THREE TYPES OF LIBs [33]

| Types | LFP | NMC | NCA |
|---|---|---|---|
| $E_a (J/mol)$ | 30336 | 28775 | 2992 |
| $\beta_0$ | 77.86 | 128.29 | 0.000061 |
| $\lambda_0 (m^{-1})$ | 0.83 | 0.33 | -120838 |
| $m$ | 2.60 | 2.62 | 1.17 |
| Nominal Capacity $C_n (Ah)$ | 2.30 | 1.50 | 0.40 |
| $S_{neg} (m^2)$ | 7.76 | 1.17 | 15.56 |

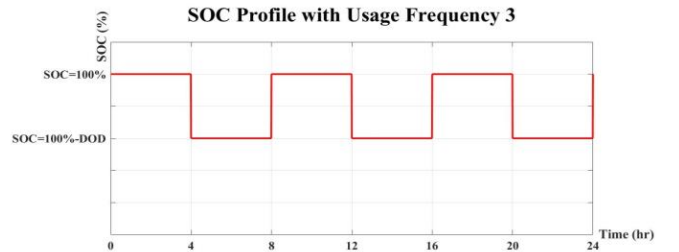

Fig. 1. An Example of SOC Profile of the Batteries with $f = 3$

TABLE II
CALENDAR AGING MODEL PARAMETERS FOR THE THREE TYPES OF LIBS [42]

| Types | LFP | NMC | NCA |
|---|---|---|---|
| $\alpha_1$ | $1.375 \times 10^{-4}$ | $1.4 \times 10^{-4}$ | $1.3 \times 10^{-4}$ |
| $\alpha_2$ | $5.5 \times 10^{-5}$ | $4.8 \times 10^{-5}$ | $5.0 \times 10^{-5}$ |
| $\alpha_3$ | $1.1 \times 10^{-4}$ | $1.0 \times 10^{-4}$ | $0.9 \times 10^{-4}$ |
| $T_0(°C)$ | 10 | 8 | 15 |

### B. Modeling the Aging Process

Battery capacity fade due to charging process can be estimated using the model introduced in Section II-A. However, battery capacity will also decrease due to self-discharge. Self-discharge rate varies with operating conditions, and severe capacity loss may be caused by extreme environment. Self-discharge rates under different operating scenarios have been discussed in [43], which can be referred when using our model in a specific scene. In this article, we use $v_1$ and $v_2$ to represent capacity fade due to usage and self-discharge rate respectively and $v_2$ is calculated by using the sum of $Q_{loss}^{SOC}$ and $Q_{loss}^{full\ SOC}$ in the following study.

In practice, battery assets cannot work continuously due to interim breakdown or environmental factors. Therefore, usage rate of battery assets $u$ is introduced into the problem setting, which means a battery asset is available during the percentage $u$ of the total working time. Considering aging effect due to usage, self-discharge, and average availability, the overall capacity loss $v$ during each time period can be described as (7), which will be incorporated in the modified parallel asset management model in Section III.

$$v = uv_1 + (1-u)v_2 \quad (7)$$

### C. Battery Lifetime Prediction

The nonlinear capacity fade model introduced in Section II-A is applied to predict battery degradation under different operating conditions. Factors considered in this paper include temperature, depth of discharge and charge rate. Table. III lists operating combinations of three factor levels and the predicted LFP battery lifetimes which are obtained based on a usage frequency of 3 cycle/day and end-of life capacity of 75%.

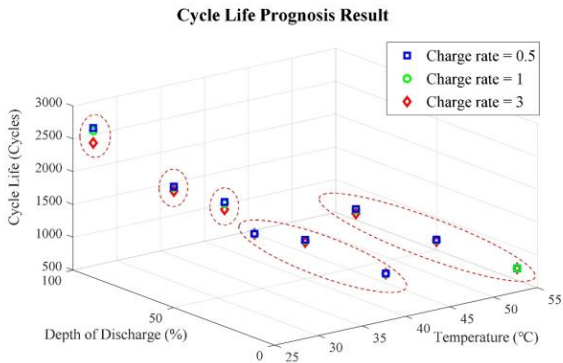

Fig. 2. Battery Cycle Life Prognosis Results

TABLE III
SCENARIO SPECIFICATIONS AND RESULTS

| Scenario | Temper-ature (°C) | DOD (%) | Charge Rate (C) | Predicted Cycle Life (Cycle) | Predicted Lifetime (Month) |
|---|---|---|---|---|---|
| 1 | 25 | 90 | 0.5 | 2780 | 31 |
| 2 | 40 | 90 | 0.5 | 1212 | 13 |
| 3 | 55 | 90 | 0.5 | 666 | 7 |
| **4** | **25** | **50** | **0.5** | **2343** | **26** |
| 5 | 40 | 50 | 0.5 | 1094 | 12 |
| 6 | 55 | 50 | 0.5 | 657 | 7 |
| 7 | 25 | 10 | 0.5 | 2085 | 23 |
| 8 | 40 | 10 | 0.5 | 1038 | 12 |
| 9 | 55 | 10 | 0.5 | 679 | 8 |
| **10** | **25** | **90** | **1** | **2738** | **30** |
| 11 | 40 | 90 | 1 | 1190 | 13 |
| 12 | 55 | 90 | 1 | 652 | 7 |
| 13 | 25 | 50 | 1 | 2330 | 26 |
| **14** | **40** | **50** | **1** | **1086** | **12** |
| 15 | 55 | 50 | 1 | 652 | 2084 |
| **16** | **25** | **10** | **1** | **2084** | **23** |
| 17 | 40 | 10 | 1 | 1037 | 12 |
| 18 | 55 | 10 | 1 | 679 | 8 |
| 19 | 25 | 90 | 3 | 2552 | 28 |
| 20 | 40 | 90 | 3 | 1099 | 12 |
| 21 | 55 | 90 | 3 | 597 | 7 |
| 22 | 25 | 50 | 3 | 2276 | 25 |
| 23 | 40 | 50 | 3 | 1059 | 12 |
| 24 | 55 | 50 | 3 | 634 | 7 |
| 25 | 25 | 10 | 3 | 2081 | 23 |
| 26 | 40 | 10 | 3 | 1036 | 12 |
| **27** | **55** | **10** | **3** | **678** | **8** |

Table. III shows that different operating conditions may lead to the same battery lifetime and similar degradation pattern. Therefore, we select 5 different scenarios marked in bold with different battery lifetimes to conduct parametric studies in Section IV. The battery cycle life prediction results and accordingly 5 different prognosis scenarios can be seen clearly in Fig. 2, in which operating conditions leading to the same lifetime result have been grouped together using circles.

### III. BATTERY ASSET MANAGEMENT MODEL

This section will introduce how battery lifetime prediction is incorporated in the parallel asset management model so that a minimum-total-cost replacement schedule for each individual asset considering battery lifetime characteristics can be determined. By solving the battery asset management model, we aim to identify the best policy of purchasing, using, storing and salvaging assets. The flow chart in Fig. 3 summarizes the battery asset management model, which is built upon both parallel asset management model and battery lifetime prediction model.

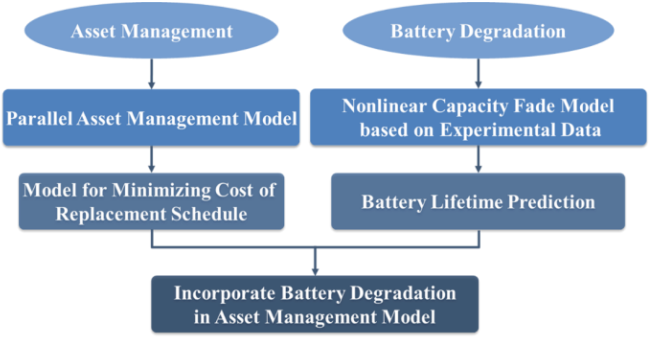

Fig. 3. Battery Degradation Model and Asset Management Model Framework

A. *Nomenclature*

**Indices:**

$i$ : index for asset age;

$j$ : index for time period or time point;

$n$ : maximum age of an asset;

$m$ : number of time periods.

**Parameters:**

$P_j$ : cost of purchasing one unit of battery at time point $j$;

$K_j$ : fixed cost of purchasing battery assets at time point $j$;

$C_j$ : operation and maintenance (O&M) cost for one unit of battery at time period $j$;

$H_j$ : inventory cost for one unit of battery at time period $j$;

$R_{ij}$ : salvage revenue from one unit of battery with age $i$ at time point $j$;

$N_i$ : number of initial batteries at age $i$;

$a$ : initial battery capacity of one unit of battery;

$d_j$ : electricity demand at time period $j$;

$v_i$ : capacity loss percentage of battery with age $i$ considering usage and self-discharge;

$u$ : usage rate of battery assets.

**Decision Variables:**

$B_j$ : number of batteries to purchase at time point $j$;

$Z_j$ : binary variable, which equals 1 if battery assets are purchased at time point $j$, and 0 otherwise;

$X_{ij}$ : number of batteries in use with age $i$ at time period $j+1$;

$I_{ij}$ : number of batteries in storage with age $i$ at time period $j+1$;

$S_{ij}$ : number of salvaged batteries with age $i$ at time point $j$.

B. *Model Assumptions*

The following model assumptions have been made in this study.

a) Operating condition is fixed over the decision horizon.
b) The capacity of a battery asset (in usage and in inventory) decreases due to usage and self-discharge after each month but is regarded invariant within each month.
c) Salvage revenue is related with asset age while O&M and inventory costs are not.
d) No battery assets are salvaged at the initial time point 0.

C. *Integer Programming Model*

Under each operating condition, periodical battery capacity degradation can be calculated with the method introduced in Section II-A, battery aging index considering usage rate can be determined as explained in Section II-B, and battery lifetime can be predicted given an end-of-life capacity threshold as shown in Section II-C. Then, the following mixed-integer optimization model can be established specifically.

$$\min \sum_{j=0}^{m-1}(P_j B_j + K_j Z_j) + \sum_{i=0}^{n-1}\sum_{j=0}^{m-1}(C_j X_{ij} + H_j I_{ij}) - \sum_{i=1}^{n}\sum_{j=1}^{m} R_{ij} S_{ij} \quad (8)$$

*Subject to*

$$\sum_{i=0}^{n-1}(1-v_i)au X_{ij} \geq d_j, \forall j = 0,...,m-1 \quad (9)$$

$$X_{i0} + I_{i0} = N_i, \forall i = 1,...,n-1 \quad (10)$$

$$X_{00} + I_{00} - B_0 = N_0 \quad (11)$$

$$X_{ij} + I_{ij} + S_{ij} - X_{(i-1)(j-1)} - I_{(i-1)(j-1)} = 0, \forall i=1,...,n-1, j=1,...,m-1 \quad (12)$$

$$S_{nj} - X_{(n-1)(j-1)} - I_{(n-1)(j-1)} = 0, \forall j=1,...,m-1 \quad (13)$$

$$S_{im} - X_{(i-1)(m-1)} - I_{(i-1)(m-1)} = 0, \forall i=1,...,n \quad (14)$$

$$X_{0j} + I_{0j} - B_j = 0, \forall j=1,...,m-1 \quad (15)$$

$$B_j \leq \{d_j / [(1-v_n)ua]\} Z_j, \forall j=0,...,m-1 \quad (16)$$

$$X_{ij}, I_{ij} \in \{0,1,2,...\}, \forall i=0,...,n-1, j=0,...,m-1 \quad (17)$$

$$S_{ij} \in \{0,1,2,...\}, \forall i=1,...,n, j=1,...,m \quad (18)$$

$$B_j \in \{0,1,2,...\}, \forall j=0,...,m-1 \quad (19)$$

$$Z_j \in \{0,1\}, \forall j=0,...,m-1 \quad (20)$$

The objective function (8) minimizes the cost of purchase, O&M and inventory minus the revenue from salvaged assets. Constraint (9) guarantees that electricity demand is satisfied at each time period. Available capacity at a certain time period is calculated considering aging index due to usage and self-discharge rate as introduced in Section II-B. Constraint (10) and (11) describe initial condition of the system: battery assets that the system has already had can be either used or stored and new assets in the system should be assigned together with initial purchase. Flow is conserved through the constraint (12). Constraints (13) and (14) describe the final condition of the system: assets at the maximum age should be salvaged and all assets need to be salvaged at the end of the decision horizon. Constraint (15) ensures that newly-purchased assets should be either used or stored at the time period of purchase. Constraint (16) enforces whenever any assets are purchased at any time point, a fixed cost will be incurred in the objective function. Finally, constraints (17)-(20) define the range of integer variables.

D. *Optimal Asset Management Schedule*

With the integer programming model built, an optimal policy for battery asset management under certain operation scenarios can be found. We consider a quarter as a time period and 75% of the original capacity as the end-of-life capacity. With the nonlinear capacity fade model in Section II-A and aging index calculation formula in Section II-B, battery capacity degradation can be obtained given a certain operating condition. Quarterly demand is calculated based on the monthly electricity total retail sales data from 2014 to 2018 in US. The demand input in this study is 1/1000 of the original data so that the demand scale fits the parameter setting well. We first use battery assets with 3-quarter lifetime and solve the model under the parameter setting in Table. IV to find characteristics of the optimal policy. In the table, inflation rate is used to calculate the periodical cost and revenue based on each initial value, i.e. $P_2 = P_0 \cdot (1+r)^2$.

TABLE IV
EXPERIMENT SETTING

| Model Input | Symbol | Value |
| --- | --- | --- |
| Usage frequency of battery assets | $f$ | 3 cycles/day |
| End-of-life capacity | $C_{EOL}$ | 75% of the total capacity |
| Number of time periods (quarters) | $m$ | 20 |
| Usage rate | $u$ | 80% |
| Initial capacity of one unit of battery | $a$ | 8100 kWh |
| Demand for each time period | $d_j$ | 940,572,000 kWh on average |
| Unit purchase cost at time point 0 | $P_0$ | 250 $ |
| Purchase fixed cost at time point 0 | $K_0$ | 40 $ |
| Operation and maintenance cost at time point 0 | $C_0$ | 10 $ |
| Inventory cost at time point 0 | $H_0$ | 10 $ |
| Salvage revenue of new assets at time point 1 | $R_{11}$ | 20 $ |
| Inflation rate (quarterly) | $r$ | 0.24% |
| Number of initial new battery assets | $N_0$ | 120 |

When battery assets with 3-quarter lifetime are applied to the energy storage system, the operation schedule obtained from the proposed model with parametric setting listed in Table. IV is shown in Fig. 4. Most of the assets are utilized till their end of life and then salvaged except 1 salvaged after one-period usage and 7 salvaged after two-period usage before the last time point. Since redundant assets can always be salvaged instead of being stored to save operation cost, the number of assets that are put into inventory is always zero in this case. And the two lines representing demand and actual capacity provided by the battery assets almost coincides due to the purpose of satisfying periodical demand with the minimum operating cost.

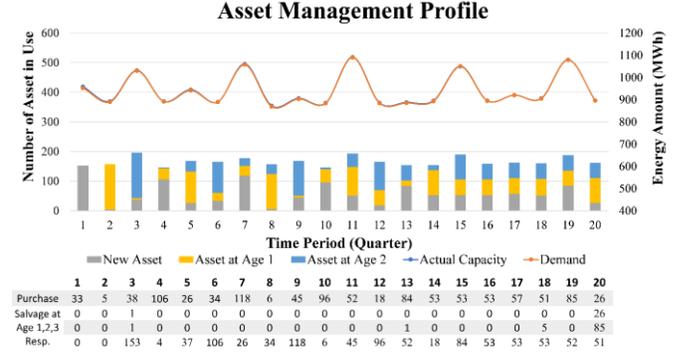

Fig. 4. An Example of Operation Schedule

Since we have obtained 5 different lengths of asset usage in Table. III, we use battery assets with 5 different lifetimes to solve for optimal management schedules and obtain results in Table. V. There are 20 time periods in the study corresponding to 21 time points numbered from 0 to 20. Since all assets will be salvaged at the last time point to minimize the total cost, the average asset salvage age and salvage time points are displayed based on the result from time point 0 to time point 19. We compare the minimum total cost from the optimal solution with the operation cost calculated from a simple heuristic, in which assets are always utilized till the end-of life and periodical purchase decision is based on the gap between remaining capacity and the quarterly demand. From the comparison, it is clear that the optimal management schedule can bring cost savings. When battery lifetime varies, the time points that salvage or purchase has to be performed will also change but assets tend to be salvaged near their end-of life. However, there are indeed cases that cannot be detected intuitively when salvaging assets earlier will lead to a long-term benefit, which is also the advantage of using the mathematical programming model proposed in this paper.

TABLE V
ASSET MANAGEMENT DECISIONS

| | Average Salvage Age | Salvage Time Points | Purchase Time Points | Optimal Cost ($) | Cost based on Heuristic ($) |
| --- | --- | --- | --- | --- | --- |
| $L=3$ | 2.99 | 3,18 | All | 280578.11 | 280635.95 |
| $L=4$ | 3.83 | 3,5,7,11,12,13,15,16,19 | 0,2,4,5,6,8,9,10,12,13,14,16,17,18 | 218988.22 | 220210.20 |
| $L=8$ | 7.39 | 3,5,7,11,15,19 | 0,2,6,8,10,13,14,16,17,18 | 131732.42 | 133557.83 |
| $L=9$ | 8.55 | 3,5,7,11,14,15,16,17,18 | 0,2,6,8,9,10,13,14,16,17,18 | 120391.48 | 121175.71 |
| $L=10$ | 9.22 | 7,10,11,15,19 | 0,2,6,10,13,14,18 | 98669.02 | 100222.48 |

IV. PARAMETRIC STUDIES

Different operating scenarios and market situation will influence the decision-making process to varying degrees. In this section, parametric studies are conducted to evaluate the influence of different model inputs for a five-year time horizon. We suppose there are only new battery assets and the number of assets at other ages equals to 0 at the initial condition. Parameters in this model and their relationship are summarized in Fig. 5, in which an influence factor is linked to an affected factor with an arrow. The unit purchase cost and salvage revenue for batteries with larger capacity will be higher than that for ones with smaller capacity. Unit battery capacity, operating condition, and usage frequency will influence the capacity degradation due to usage based on the method introduced in Section II-A. Meanwhile, operating condition and usage frequency will also influence the self-discharge rate. Then, the capacity degradation and self-discharge, together with usage rate will determine the aging index. When exploring the effect of a certain parameter, we change the related values while keeping others same as listed in Table IV.

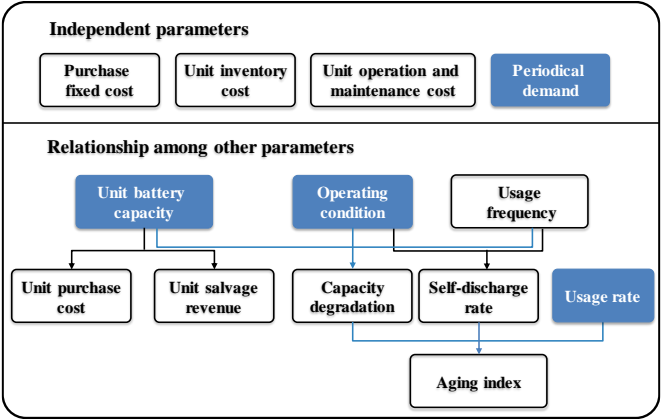

Fig. 5. Parameters and Their Relationship

### A. *Effect of Usage Rate*

In this section, we explore the effect of usage rate of the battery assets on the system performance. Usage rate is introduced in this model to accommodate emergencies and provide flexibility in the usage formula of assets as well. We change the usage rate from 10% to 90% under 5 different operating conditions marked in Table. III in the experiment, which leads to the result in Fig. 6.

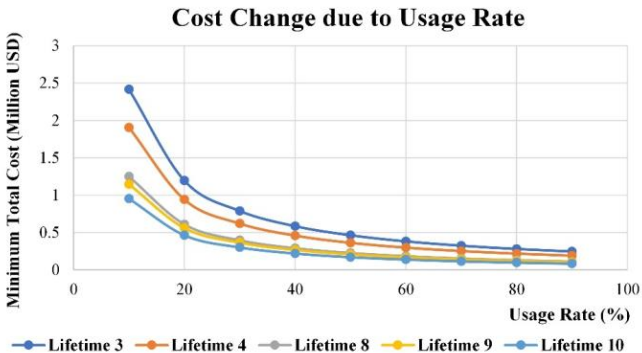

Fig. 6. Cost Change due to Usage Rate

Fig. 6 shows that the minimum total cost for a five-year time horizon decreases as the usage rate increases. Difference between operating conditions is reflected in lifetime since operating conditions leading to the same predicted lifetime are considered equivalent. And scenarios with different battery lifetimes present similar decreasing pattern as battery assets are utilized for a larger percentage of time. The reason is that when usage rate is higher, the system needs smaller number of battery assets to satisfy the demand which will reduce purchase cost. As other types of cost remain the same level, the minimum total cost will finally decrease.

### B. *Effect of Unit Battery Capacity*

Effect of unit battery capacity is discussed in this section since battery assets with various performance indicators are available in the market. Should the company always purchase batteries with the highest capacity? The answer to this question may lead to a valuable decision guidance. As the increase of battery capacity, its price will also rise and the depreciation process may fluctuate. In this section, we suppose unit purchase cost and salvage revenue increase proportionally as the battery capacity is augmented while purchase fixed cost, maintenance and inventory cost remain invariant. At each run, we change the number of initial new battery assets so that demand for the first time period can be satisfied at full utilization. By using battery with 9-quarter lifetime and changing unit battery capacity from 3600 *kWh* to 18000 *kWh* , we can acquire the result in Fig. 7.

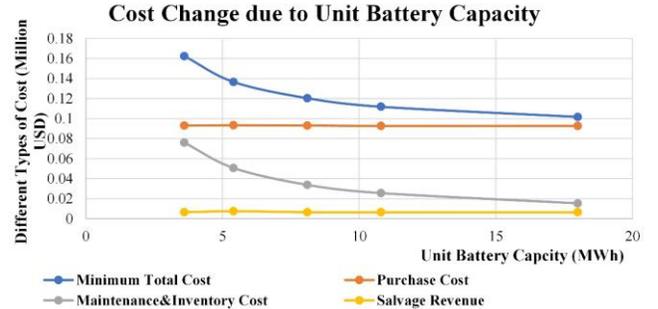

Fig. 7. Cost Change due to Unit Battery Capacity

We are informed from Fig. 7 that the minimum total operating cost will decrease as the unit battery capacity increases. The reason is that the number of battery assets we need to satisfy periodical demand will be smaller. Since O&M cost is only related with the number of assets, it will decrease under this condition. And as the increment of battery capacity, unit purchase cost and salvage revenue increase proportionally, so the total purchase cost and salvage revenue will almost remain the same. Therefore, the total operating cost decreases as the unit battery capacity increases mainly due to the reduction of maintenance cost.

We can also notice from Fig. 7 that the decrement of total operating cost turns insignificant as the unit battery capacity increases to a certain amount, which provides us an insight in battery selection. Companies may not have to pursue extremely large capacity of battery assets since a reasonable amount is enough for the low-cost operation purpose.

## C. Effect of Operating Condition

Battery lifetime may vary under different operating conditions. In this section, we test the effect of operating conditions, namely how battery lifetime has influenced our decision-making process. Five different operating conditions same as those in Section IV-A are selected in this experiment and the result is shown in Fig. 8.

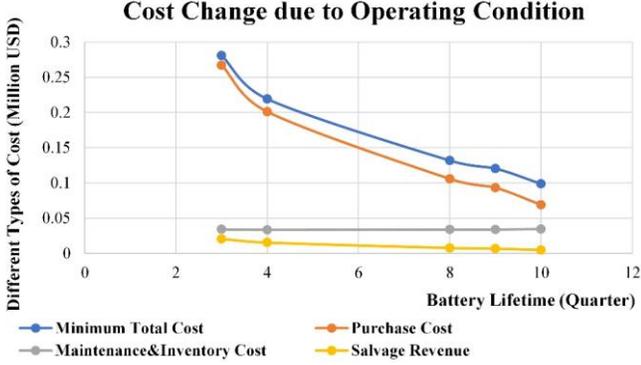
Fig. 8. Cost Change due to Operating Condition

We may notice from the result that minimum total cost decreases as the battery lifetime increases. That's because when battery assets have a longer lifetime, purchase of new assets will be at a lower frequency. Replacement decisions from the optimal solution tend to keep using old assets until they reach their maximum age since their capacity reduction is small so that the optimization model tends to use them rather than purchasing new assets. The result emphasizes the importance of keeping a moderate operating condition for the battery assets so that longer lifetime holds.

## D. Effect of Periodical Demand

We use the data of monthly electricity total retail sales from 2014 to 2018 in US to calculate the periodical demand in previous experiments. In this section, data over the 50-year horizon has been applied to explore the unit cost change due to periodical demand. In the 50-year horizon, electricity demand has been increasing as shown in Fig. 9. We select five different 20-quarter time periods in this section and compare the unit demand cost which is calculated by using the minimum total cost divided by total demand over the decision horizon to gain insights in the asset replacement decisions. The result is presented in Fig. 10.

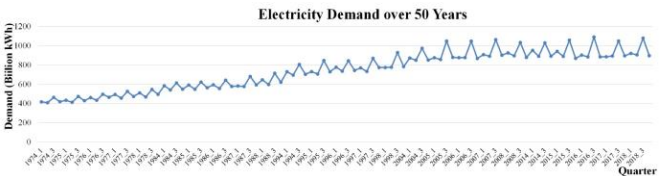
Fig. 9. Electricity Demand over 50 Years

As time goes from 1974 to 2018, unit demand cost slightly increases as the periodical demand increases with battery assets of different lifetime used. The unit demand cost is a balance between the average unit purchase price, average maintenance and inventory cost, and average salvage revenue. From the slight increase, we are informed that as demand increases the benefit from unit salvage plays a weaker role than that of all types of cost so that the unit demand cost has a slight increase. There is also a gap between the results of lifetime 3,4 and results of lifetime 8,9,10, which also illustrates the importance of maintaining an appropriate lifetime.

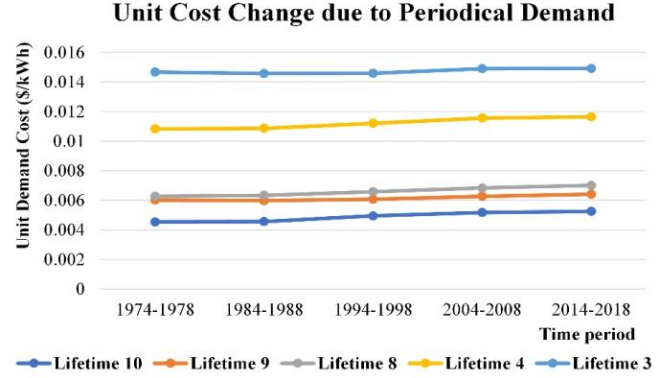
Fig. 10. Unit Cost Change due to Periodical Demand

## V. DISCUSSION

In previous sections, we regard five years as our decision horizon and a quarter as one decision period, in which process we suppose battery capacity remains the same within one month and decreases at a certain aging index and self-discharge rate after each month. However, decision horizon may vary from several months to tens of years in practice. So first we will illustrate how to apply the model proposed in this article to different decision horizons. Since extreme cases may occur in real practice, we will then discuss the asset management decisions under possible extreme cases.

When decision horizons and periods change, we still update battery capacity monthly but adjust our decision variables according to the actual decision interval. We discretize battery capacity change in a way illustrated by Fig. 11, in which the battery capacity is updated after one-month usage. When we make decisions quarterly, we use the average capacity within the quarter as the invariant periodical capacity. For example, $1-v_0 = (c_0 + c_1 + c_2)/3$ and the value for $c_0, c_1, c_2$ are calculated using the degradation model described in Section II. When we change the decision-making interval, two basic modifications should be made. First, the periodical invariant capacity should be modified based on the number of months within one decision interval. Second, inflation rate to update periodical costs and revenue also need to be altered based on current decision period. After these two modifications, our model can still provide replacement guidance for the battery asset management problem.

While the presented study considers different scenarios in the battery asset operation, there are extreme cases as discussed blow. First, when battery assets are being operated in a hostile environment or the decision period is relatively long, one extreme condition could happen is that all battery assets would reach their end of lives at each asset management decision point. Under this extreme case, the asset management policy could become straightforward since at each time point new assets should be purchased and all used asset should be salvaged. In the opposite scenario when battery assets can survive for the whole decision horizon, no purchase decision has to be made during the operation process and the assets should be salvaged only at the end. When there is a limit for the number of batteries that can be purchased or there is an upper bound for the capacity that the system can provide, the periodical demand may not always be satisfied, in which case the model proposed in this article will not be suitable. If such a satisfaction gap is allowed, then a penalty function should be added to the objective function and the demand satisfaction constraint can be eliminated. When unit battery cost varies not only along time but also due to technology improvement, the purchase cost should be changed from one single parameter to a parameter array saving costs for different periods based on market condition or prediction. Also, current model is fed with fixed demand for each time period so that an operation policy can be determined. However, when periodical demand is unknown, a learning algorithm should be used to provide sufficient capacity based on the consequences of previous decisions. And since the demand pattern is a given input of the optimization model, inventory decisions consist a negligible part in all the management decisions so that battery degradation process is treated identically for assets in usage and in inventory.

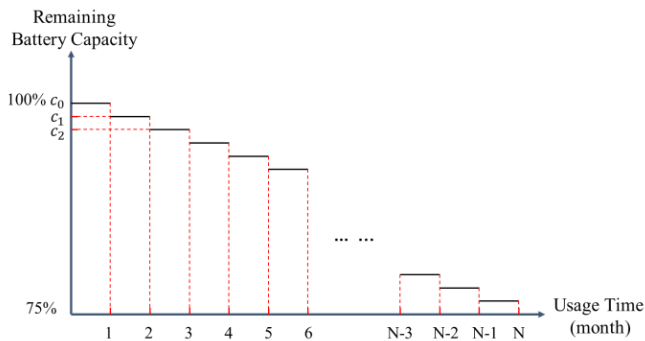

Fig. 11. Discretization of the Battery Capacity Degradation

## VI. Conclusion

We have incorporated battery lifetime prediction, which regards temperature, depth of discharge, charge rate and usage frequency as stress factors, in a parallel asset management model. Battery aging and availability are considered in the proposed model so that the decisions will be applicable to practical battery asset replacement problems. The asset management profile in Section III illustrates that by considering battery cycle life prediction in the asset management model, there is more flexibility in purchase and salvage decisions so that system owners can maximize the usage value of the purchased battery assets. Parametric studies show that an appropriate operating condition which guarantees a long battery lifetime can reduce lifecycle cost of the system and further increase in battery capacity after a certain value will lead to slight reduction in total cost. Also, an extension of the model is discussed in Section V to show how the proposed model can be applied to different decision horizons and periods.

As the future work, profits from the battery energy storage system should be introduced to the model and a continuous way of depicting battery aging process may be established. Since utility companies make profits by charging the batteries at low electricity price and discharge at high price, the asset management model should also incorporate the fluctuation of electricity price together with demand. And BESS in real practice may consist of mixed types of batteries with varying degradation characteristics and capacity which also requires further research. Finally, battery asset management profile may also change if budget limit occurs in the decision-making process.

REFERENCES

[1] G. P. Kulkarni *et al.,* "The united states of storage [electric energy storage]," *IEEE Power and Energy Magazine 3,* no. 2, pp. 31-39, Mar. 2005.

[2] K. C. Divya, and J. Østergaard, "Battery energy storage technology for power systems—An overview," *ELECTR POW SYST RES.*, vol. 79, no. 4, pp. 511-520, Apr. 2009.

[3] B. Dunn, H. Kamath, and J.-M. Tarascon, "Electrical Energy Storage for the Grid: A Battery of Choices," *Science*, vol. 334, no. 6058, pp. 928–935, Nov. 2011.

[4] A. Khaligh and Z. Li, "Battery, Ultracapacitor, Fuel Cell, and Hybrid Energy Storage Systems for Electric, Hybrid Electric, Fuel Cell, and Plug-In Hybrid Electric Vehicles: State of the Art," *IEEE Transactions on Vehicular Technology*, vol. 59, no. 6, pp. 2806–2814, Apr. 2010.

[5] L. Lu *et al.,* "A review on the key issues for lithium-ion battery management in electric vehicles," *J POWER SOURCES.,* vol. 226, pp. 272-288, Mar. 2013.

[6] M. V. Berecibar *et al.,* "Critical review of state of health estimation methods of Li-ion batteries for real applications," *RENEW SUST ENERG REV.,* vol. 56, pp. 572-587, Apr. 2016.

[7] K. B. Hatzell, A. Sharma, and H. K. Fathy. (2012, June). A survey of long-term health modeling, estimation, and control of Lithium-ion batteries: Challenges and opportunities. Presented at ACC. [Online]. Available: https://ieeexplore.ieee.org/abstract/document/6315578?casa_token=2Ujdlgi4v_QAAAAA:YZd8QK1OR9XtRvpwC-nG9TnPKkEcqkxg0-imhT91BB7BdflJAbvk-mw5DKSvao8Tpspwgs9rTA

[8] S. Piller, M. Perrin, and A. Jossen, "Methods for state-of-charge determination and their applications," *J Power Sources.,* vol. 96, no. 1, pp. 113–120, Jun. 2001.

[9] J. Zhang and J. Lee, "A review on prognostics and health monitoring of Li-ion battery," *J Power Sources.,* vol. 196, no. 15, pp. 6007–6014, Aug. 2011.

[10] System Monitoring the Discharging Period of the Charging/Discharging Cycles of Rechargeable Battery, and Host Device Including a Smart Battery, by J.N. Patillon. (1999, Aug. 10) Patent 5,936,385 [Online]. Available: https://patentimages.storage.googleapis.com/0c/7c/33/8b4b6e4925fc60/US5936385.pdf

[11] C. Chan, E.W.C. Lo, and W. Shen, "The available capacity computation model based on artificial neural network for lead–acid batteries in electric vehicles," *J Power Sources.,* vol. 87, no. 1-2, pp. 201–204, Apr. 2000.

[12] F. Heut, "A review of impedance measurement for determination of state-of-charge or state-of-health of secondary battery," *J Power Sources.,* vol. 70, pp. 59-69, 1998.

[13] S. Rodrigues, N. Munichandraiah, and A.K. Shukla, "A review of state-of-charge indication of batteries by means of ac impedance measurements," *J Power Sources.*, vol. 87, no. 1-2, pp. 12-20, Apr. 2000.

[14] A.J. Salkind, C. Fennie, and P. Singh, "Determination of state-of-charge and state-of-health of batteries by fuzzy logic methodology," *J Power Sources.*, vol. 80, no. 1-2, pp. 293-300, Jul. 1999.

[15] J. Kozlowski, (2003, March). Electrochemical cell prognostics using online impedance measurements and model-based data fusion techniques. Presented at AeroConf. [Online]. Available: https://ieeexplore.ieee.org/abstract/document/1234169?casa_token=4_Al3qFnLI8AAAAA:WacgMNC_hkGEaxW-O02CJdW8bFo806gi1i7-sl4oM3NGlJBKSyU6-zg2CRzvXZWzMaaxr_HmZw

[16] B. Saha et al., "Prognostics methods for battery health monitoring using a Bayesian framework," *IEEE T INSTRUM MEAS.*, vol. 58, no. 2, pp. 291-296, Oct. 2009.

[17] W. He et al., "Prognostics of lithium-ion batteries based on Dempster–Shafer theory and the Bayesian Monte Carlo method," *J Power Sources.*, vol. 196, no. 23, pp. 10314-10321, Dec. 2011.

[18] G. Bai, P. Wang, and C. Hu, "A self-cognizant dynamic system approach for prognostics and health management," *J Power Sources.*, vol. 278, pp. 163-174, Mar. 2015.

[19] C. Hu et al., (2014, June). Method for estimating capacity and predicting remaining useful life of lithium-ion battery. Presented at PHM. [Online]. Available: https://ieeexplore.ieee.org/abstract/document/7036362?casa_token=od02SFiuZ2kAAAAA:dNwLtN-UfXcAJSxrtCed0uec_3wG22Q5UECnQJ6LsDZBvmopVsNcD99QeyH_t8abHYNiM5Knpw

[20] F. Sun et al., "Model-based dynamic multi-parameter method for peak power estimation of lithium–ion batteries," *APPL ENERG.*, vol. 96, pp. 378-386, Aug. 2012.

[21] W. Waag, S. Käbitz, and D.U. Sauer, "Experimental investigation of the lithium-ion battery impedance characteristic at various conditions and aging states and its influence on the application," *APPL ENERG.*, vol. 102, pp. 885-897, Feb. 2013.

[22] JE. Amadi-Echendu et al., "What Is Engineering Asset Management?" in *Definitions, Concepts and Scope of Engineering Asset Management* London, U.K. Springer, 2010, pp. 3-16.

[23] K. El-Akruti, R. Dwight, T. Zhang, "The strategic role of engineering asset management," *INT J PROD ECON.*, vol. 146, no. 1, pp. 227-239, Nov. 2013.

[24] Y. Yatsenko, and N. Hritonenko, "Asset replacement under improving operating and capital costs: a practical approach," *INT J PROD RES.*, vol. 54, no. 10, pp. 2922-2933, May 2016.

[25] E. des-Bordes, and İ.E.Büyüktahtakın, "Optimizing capital investments under technological change and deterioration: A case study on MRI machine replacement," *ENG ECON.*, vol. 62, no. 2, pp. 105-131, Apr. 2017.

[26] İ.E.Büyüktahtakın et al., "Parallel asset replacement problem under economies of scale with multiple challengers," *ENG ECON.*, vol. 59, no. 4, pp. 237-258, Oct. 2014.

[27] İ.E.Büyüktahtakın and J.C. Hartman, "A mixed-integer programming approach to the parallel replacement problem under technological change," *INT J PROD RES.*, vol. 54, no. 3, pp. 680-695, Feb. 2016.

[28] N. Hritonenko, and Y. Yatsenko, "Fleet replacement under technological shocks," *ANN OPER RES.*, vol. 196, no. 1, pp. 311-331, Jul. 2012.

[29] J.L. Rogers, and J.C. Hartman, "Equipment replacement under continuous and discontinuous technological change," *IMA J Manag Math.*, vol. 16, no. 1, pp. 23-36, Jan. 2005.

[30] M. Safari et al., "Multimodal physics-based aging model for life prediction of Li-ion batteries," *J ELECTROCHEM SOC.*, vol. 156, no. 3, pp. A145-A153, Dec. 2008.

[31] Y. Cui et al., "Multi-stress factor model for cycle lifetime prediction of lithium ion batteries with shallow-depth discharge," *J Power Sources.*, vol. 279, pp. 123-132, Apr. 2015.

[32] R. Rao, S. Vrudhula, and D.N. Rakhmatov, "Battery modeling for energy aware system design," *COMPUTER,* vol. 36, no. 12, pp. 77-87, Dec. 2003.

[33] J.F. Li et al., "Three-parameter modeling of nonlinear capacity fade for Lithium-Ion batteries at various cycling conditions," *J ELECTROCHEM SOC.*, vol. 164, no. 12, pp. A2767-A2776, Sep. 2017.

[34] X. Lin et al., "A comprehensive capacity fade model and analysis for Li-ion batteries," *J ELECTROCHEM SOC.*, vol. 160, no. 10, pp. A1701-A1710, Aug. 2013.

[35] G. Ning, R.E. White, and B.N. Popov, "A generalized cycle life model of rechargeable Li-ion batteries," *ELECTROCHIM ACTA.*, vol. 51, no. 10, pp. 2012-2022, Feb. 2006.

[36] M. Safari et al., "Multimodal physics-based aging model for life prediction of Li-ion batteries," *J ELECTROCHEM SOC.*, vol. 156, no. 3, pp. A145-A153, Dec. 2008.

[37] A.A. Tahmasbi, T. Kadyk, and M.H. Eikerling, "Statistical physics-based model of solid electrolyte interphase growth in lithium ion batteries," *J ELECTROCHEM SOC.*, vol. 164, no. 6, pp. A1307-A1313, Apr. 2017.

[38] R. Fu et al., "Development of a physics-based degradation model for lithium-ion polymer batteries considering side reactions," *J Power Sources.*, vol. 278, pp. 506-521, Mar. 2015.

[39] S.S. Choi, and H.S. Lim, "Factors that affect cycle-life and possible degradation mechanisms of a Li-ion cell based on LiCoO2," *J Power Sources.*, vol. 111, no. 1, pp. 130-136, Sep. 2002.

[40] S. Kim et al., "Multiphysics coupling in lithium-ion batteries with reconstructed porous microstructures," *J PHYS CHEM C.*, vol. 122, no. 10, pp. 5280-5290, Feb. 2018.

[41] G. Fan et al., "A reduced-order multi-scale, multi-dimensional model for performance prediction of large-format li-ion cells," *J ELECTROCHEM SOC.*, vol. 164, no. 2, pp. A252-A264, Dec. 2016.

[42] P. Keil et al., "Calendar aging of lithium-ion batteries," *J ELECTROCHEM SOC.*, vol. 163, no. 9, pp. A1872-A1880, Jul. 2016.

[43] A.H. Zimmerman, "Self-discharge losses in lithium-ion cells," *IEEE AERO EL SYS MAG.*, vol. 19, no. 2, pp. 19-24, Mar. 2004.

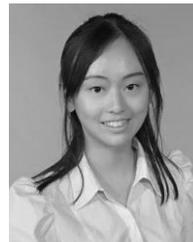

**Xinyang Liu** received the B.S. degree in industrial engineering from Tsinghua University, Beijing, China, in 2018. She received the M.S. degree in industrial engineering from University of Illinois at Urbana-Champaign, IL, USA in 2020 and she is currently pursuing the Ph.D. degree there. Her research interests include maintenance planning and data-driven decision making.

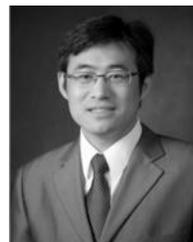

**Pingfeng Wang** (M'09) received the B.S. degree in mechanical engineering from University of Science and Technology, Beijing, China, in 2001, the M.S. degree in applied mathematics from Tsinghua University, Beijing, China, in 2006, and the Ph.D. degree in mechanical engineering from University of Maryland, College Park, MD, USA, in 2010.

He is currently the Jerry S. Dobrovolny Faculty Scholar and an Associate Professor in the Department of Industrial and Enterprise Systems Engineering, University of Illinois at Urbana-Champaign, Urbana, IL, USA. His dedicated research efforts have resulted in more than 100 publications in refereed journals and conference proceedings. His research interests include engineering system design for reliability failure resilience and sustainability, and prognostics and health management.

Dr. Wang's research has garnished him notable international awards including two times ASME Best Paper Awards in 2008 and 2013, respectively. 2012 IEEE PHM Best Paper Award, the National Science Foundation CAREER Award in 2014, the Young Researcher Award from the International Society of Green Manufacturing and Applications in 2012, and the Design Automation Young Investigator Award from the American Society of Mechanical Engineers in 2016.

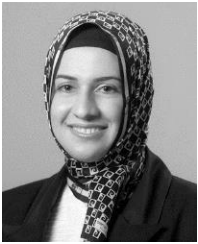

**Esra Büyüktahtakın Toy** received the B.S. degree in industrial engineering from Fatih University, Turkey, in 2002, the M.S. degree in industrial engineering from Bilkent University in 2005, the M.S. degree in Management Science from Lehigh University in 2007, and the Ph.D. degree in industrial and systems engineering from the University of Florida in 2009.

She is currently an Associate Professor in the Department of Mechanical and Industrial Engineering, New Jersey Institute of Technology. Her research interests focus on optimization, mixed-integer programming, and data analytics with non-traditional applications of OR in environment, sustainability, and health-care.

Dr. Buyuktahtakin Toy is the recipient of the NSF CAREER Award co-funded by the ENG Environmental Sustainability program and the NSF Division of Mathematical Sciences (Applied Mathematics, Computational Mathematics, and Mathematical Biology programs).

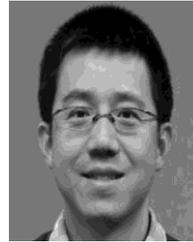

**Zhi Zhou** (M'09) received the B.Eng. and M.Eng. degrees in computer science from Wuhan University, Wuhan, China, in 2001 and 2004, respectively, and the Ph.D. degree in decision sciences and engineering systems from Rensselaer Polytechnic Institute, Troy, NY, USA, in 2010.

He is currently a Computational Engineer in the Energy Systems Division at Argonne National Laboratory, Lemont, IL, USA. His research interests include agent-based modeling and simulation, stochastic optimization, statistical forecasting, electricity markets, and renewable energy.